\documentclass[12pt,preprint]{aastex}

\def\cm{{\rm\thinspace cm}}
\def\erg{{\rm\thinspace erg}}
\def\s{{\rm\thinspace s}}
\def\ergps{\hbox{$\erg\s^{-1}\,$}}
\def\pcmsq{\hbox{$\cm^{-2}\,$}}

\begin{document}

\title{Chandra Observations of Nuclear X-ray Emission from Low Surface 
Brightness Galaxies}

\author{M.~Das\altaffilmark{1}, C.S.~Reynolds\altaffilmark{2},
S.N.~Vogel\altaffilmark{2}, S.S.~McGaugh\altaffilmark{2},
N.G.~Kantharia\altaffilmark{3}}

\affil{\altaffilmark{1}Raman Research Institute, Sadashivanagar, Bangalore, India}
\affil{\altaffilmark{2}Department of Astronomy, University of Maryland,
College Park, MD 20742, USA}
\affil{\altaffilmark{3}National Center for Radio Astrophysics, Tata Institute of Fundamental Research, 
Post Bag 3,  Ganeshkhind, Pune - 411007, India}

\email{email : mousumi@rri.res.in}

\date{Accepted.....; Received .....}

\begin{abstract}
We present Chandra detections of x-ray emission from the AGN in two
giant Low Surface Brightness (LSB) galaxies, UGC~2936 and
UGC~1455. Their x-ray luminosities are $1.8\times10^{42}~ergs~s^{-1}$
and $1.1\times10^{40}~ergs~s^{-1}$ respectively. Of the two galaxies,
UGC~2936 is radio loud.  Together with another LSB galaxy UGC~6614
(XMM archival data) both appear to lie above the X-ray-Radio fundamental
plane and their AGN have black hole masses that are low compared to
similar galaxies lying on the correlation. However, the bulges in
these galaxies are well developed and we detect diffuse x-ray emission
from four of the eight galaxies in our sample. Our results suggest that the
bulges of giant LSB galaxies evolve independently of their halo
dominated disks which are low in star
formation and disk dynamics. The centers follow an evolutionary path similar to
that of bulge dominated normal galaxies on the Hubble Sequence but 
the LSB disks remain unevolved. Thus the bulge and disk evolution 
are decoupled and so whatever star formation processes
produced the bulges did not affect the disks.
\end{abstract}


\bigskip

\bigskip

\section{Introduction}

Low Surface Brightness (LSB) galaxies are poorly evolved systems that
have diffuse stellar disks, large HI disks and massive dark halos 
\citep{ImpeyBothun1997}. The dark halo inhibits the growth of disk
instabilities such as bars and spiral arms, this leads to an overall
poor star formation rate over these galaxies
\citep{Boissier.etal.2008,O'Neil.etal.2007}. As a result LSB galaxies
are metal poor \citep{deNaray.etal.2004} and optically dim. Optical
studies show that they span a wide range of morphologies from dwarfs
to giant spirals \citep{Beijersbergen.etal.1999}. Our paper focuses on
giant LSB spirals which are
characterised by prominent central bulges, optically dim disks and
extended HI gas disks, a good example being Malin~1 
\citep{Barth.2007,ImpeyBothun1989}. These galaxies are also fairly
isolated systems \citep{Rosenbaum.Bomans}. The origin and evolution of these
galaxies is still unclear; one possibility is that they form in low 
density enviroments and hence remain unevolved \citep{Hoffman.etal.1992}.

Although LSB galaxy disks have been studied at length at optical and
infrared wavelengths
\citep{Burkholder.etal.2001,Rahman.etal.2007,Hinz.etal.2007} not much
is known about their nuclear properties. In particular, their nuclear
black hole (BH) masses and AGN-bulge evolution history remain largely
unconstrained. Optical spectroscopy indicates that a significant
fraction have Active Galactic Nuclei (AGN)
\citep{Sprayberry.etal.1995} and large bulges \citep{Schombert.1998}.
Nuclear acivity has also been detected at radio wavelengths (Das et
al. 2008, in preparation). However, the best way to detect and study
AGN activity is using x-ray emission which will give rise to a compact
x-ray bright source at the galaxy nucleus. There may also be a
primordial x-ray emitting gaseous halo which will appear as diffuse
x-ray emission associated with the galaxy or its bulge. Very little is
known about BH masses in bulge dominated LSB galaxies but indirect
studies suggest that it is low compared to normal galaxies
\citep{Pizzella.etal.2005}. This raises interesting questions
regarding the evolution of BHs, AGN and bulges in poorly evolved and
isolated galaxies.  In this paper we examine these issues with
observations made using the Chandra X-ray Observatory.  Essentially
nothing is known about the x-ray properties of LSB galaxies (possibly
because the lack of star formation and hence the low number of
high-mass x-ray binaries suggests low x-ray emission).  In the
following sections we present the results of a pilot study of eight
giant LSB galaxies with Chandra and discuss the
implications of our observations.

\section{Observations}

\subsection{Sample Selection}

We observed a total of eight LSB galaxies with Chandra
during Cycle~8. Our sample is based on a sample of gas rich, predominantly 
low surface brightness galaxies with optical AGN defined by Schombert (1998).
We have not independently
quantified their LSB characteristics with photometric observations. Although a few
galaxies (such as UGC~12845) are clearly LSB, surface photometry is required to
establish the LSB nature of the entire sample. Our sample was also limited by the
fact that we had only 30~Ksecs of observing time. Hence only a few nearby, giant
LSB galaxies were chosen. Thus our sample is by no means complete but does give a
first look at the x-ray properties of these galaxies. All the galaxies are large,
HI rich and nearby; they are derived from the UGC catalogue 
($v_{sys}\le 15,0000~km~s^{-1}$). We chose a subset of eight nearby 
($v_{sys}\le10,000~km~s^{-1}$) optically active LSB galaxies from this sample in 
order to maximise our chances of x-ray detection with
Chandra. Thus all the eight galaxies in our sample have prominent bulges
and optically identified AGN. In the following paragraph we give a brief description
of each galaxy. We take low surface brightness galaxies to have 
central surface brightness $>~22~mag~arcsec^{-2}$, which is clearly
below the Freeman value.

\noindent
{\bf UGC~1455}~:~The galaxy has a disk B band brightness of $22.4~mag~arcsec^{-2}$ 
\citep{Graham.2003}. At the center of the LSB disk is a prominent bulge which is 
oval in shape and may represent a small bar \citep{deJong.1996}.\\
{\bf UGC~2936}~:~This is a fairly inclined galaxy with an LSB disk fainter than
$22~mag~arcsec^{-2}$ \citep{Sprayberry.etal.1995} and a very extended HI disk. 
So it falls into the LSB category but has significant star formation
over the disk which is unusual for a LSB galaxy \citep{Pickering.etal.1999}. \\
{\bf UGC~1378}~:~This galaxy is classified as a LSB galaxy by Schombert (1998); it has a prominent
bulge but diffuse disk. Deeper photometry of the galaxy in the literature is lacking.\\
{\bf UGC~1922}~:~The galaxy has a prominent nucleus but a very featureless LSB disk; appearance
is indicative of a giant LSB galaxy based on Schomberts classification. Deeper photometry of this 
galaxy is lacking. It is one of the few LSB galaxies that have been detected in CO emission 
signifying the presence of molecular gas \citep{O'Neil.Schinnerer.2003}.\\
{\bf UGC~3059}~:~This is a fairly inclined galaxy like UGC~2936 and is classified as a 
LSB galaxy by Schombert (1998). It has a prominent bulge, diffuse
stellar disk and large HI gas disk.\\
{\bf UGC~4422}~:~Also knowm as NGC~2595, this galaxy has a disk B band brightness of 
$22.14~mag~arcsec^{-2}$ \citep{Graham.2003} and so falls into the LSB galaxy category. The galaxy has a 
bright core, small bar and prominent spiral arms. However, the disk shows signs of ongoing star formation which 
is unusual for LSB galaxies.\\
{\bf UGC~11754}~:~The only LSB classification for this galaxy is by Schombert; deeper photometry is lacking 
in the literature. However, it does have a very diffuse stellar disk similar to that seen in typical LSB 
galaxies. \\
{\bf UGC~12845}~:~This galaxy is classified as a LSB galaxy by \citep{Bothun.etal.1985} and also by 
\citep{Graham.2003} who measure the disk brightness as $22.77~mag~arcsec^{-2}$. It has a prominent bulge,
faint spiral arms and a fairly diffuse stellar disk.

\subsection{Chandra Observations and data reduction}

This was a pilot study and the total time of the
observation was 30\,ksecs. The galaxies and their observation IDs are
listed in Table~1 along with the exposure time. All the observations
were performed with the Advanced CCD Imaging Spectrometer (ACIS) with
the aimpoint placed on chip-S3 of the ACIS-S array.  The data were
reduced according to the standard threads using CIAO version 3.4 and
all spectral analysis was performed with XSPEC version 12.2.1.

\section{Analysis and Results}

\subsection{Nuclear sources}

For each galaxy, we began by using the {\tt wavdetect} source
detection algorithm to search for any compact/pointlike emission
associated with the nucleus; such emission would be a signature of AGN
activity. We detected a compact source in the center of 2 galaxies in
our sample, UGC~2936 and UGC~1455.

For UGC~2936, the photon count rate was sufficient to extract a
spectrum. The spectrum was extracted using the CIAO script {\tt
psextract}, and grouped to have 15 counts per bin in order to permit
the use of $\chi^2$ statistics (Figure~1). The 0.5--10\,keV spectrum was fitted
with a model consisting of a power law and modified by photoelectric
absorption from cold line-of-sight gas. Including only the Galactic
absorption ($N_H=1.24\times 10^{21}\pcmsq$) gave a poor fit and a very
flat photon index ($\chi^2/dof=59/16$ and $\Gamma=-1.2$). The fit
improves dramatically ($\chi^2=14.2/15$) if we allow for additional
absorption, presumably associated with UGC~2936 itself; the best
fitting model has a total absorbing column
$N_H=(5.3^{+2.2}_{-1.6})\times 10^{22}\pcmsq$ and photon index
$\Gamma=1.14^{+0.72}_{-0.63}$ (90\% confidence level for 1 free
parameter is quoted).  Figure~2 shows the confidence contours on the
($N_H,\Gamma)$-plane.  While the best fitting photon index is rather
flatter than typical AGN \citep{Winter.etal.2008}, the typical
value of $\Gamma=1.8$ lies within the 90\% error range.  The source is
clearly highly absorbed, however, with a 90\% lower limit of
$3.7\times 10^{22}\pcmsq$ to the total absorbing column.  This implies
a lower-limit to the {\it intrinsic} absorbing column in UGC~2936 of
$2.5\times 10^{22}\pcmsq$, and is consistent with this sources
classification as a Seyfert-2 galaxy \citep{Veron.Veron.2001}.
Correcting for the absorption, the 0.5--10\,keV luminosity of the best
fitting model is $1.8\times 10^{42}\ergps$.  The 0.5--10\,keV
luminosity of the best fitting model with the photon index frozen at
$\Gamma=1.8$ is $2.5\times 10^{42}\ergps$.

In the case of UGC~1455 there were not enough counts to produce a
meaningful spectrum. Instead, we used the {\tt PIMMS} package to
convert the count rate derived from {\tt wavdetect} into a
0.5--10\,keV luminosity assuming a powerlaw spectrum with $\Gamma=2$
and only Galactic absorption; the derived luminosity is $1.1\times
10^{40}\ergps$.

For both of these objects, the compact emission coincides with the
2MASS galaxy centers, lending further support to the notion that this
is AGN emission.  Indeed, both of these objects have optically
identified AGN emission lines \citep{Schombert.1998}.  The
corresponding x-ray luminosity of the AGN in UGC~2936 is comparatively
high, $L_{x}(0.5-10\,{\rm kev})\sim 1.8\times 10^{42}\ergps$ (Table~1)
and comparable to the bright nearby Seyfert galaxies that show strong
optical emission lines \citep{Heckman.etal.2005}. UGC~1455 has a
comparatively lower AGN luminosity $L_{x}(2-10\,{\rm kev})\sim10^{40}\ergps$
but is still bright compared to nearby low luminosity AGNs (LLAGNs)
such as NGC~4303 which has a luminosity of
$L_{x}\sim10^{39}~erg~s^{-1}$ \citep{Jimenez.etal.2003}.

For the galaxies that did not show x-ray emission from the nucleus we
used the lowest count rates of sources in the center of the field of
view to derive upper estimates of AGN luminosities. We again used the
CIAO program PIMMS to determine the x-ray flux values from which we
derived upper limits of the x-ray luminosities.  These limits are in
the range $L_{x}(0.5-10\,{\rm kev})\sim5\times 10^{38}-5\times 10^{39}\ergps$
and are quoted on an object-by-object basis in Table~1. This
luminosity is comparable to the x-ray luminosities of nearby LLAGNs
which have $L_{x}\sim10^{39}-10^{40}\,{\rm erg}\,{\rm s}^{-1}$ (Ho et
al. 2001). Hence it is possible that the non-detections in our sample
are a result of the galaxies being LLAGNs that are at large distances
(Table~1).

\subsection{Extended emission}

We searched for diffuse x-ray emission in the galaxies. Point sources
were located using {\tt wavdetect}. Elliptical regions around the
sources were masked out and then filled based on the local background
emission using the tool {\tt dmfilth}. We finally smoothed the images
using the routine aconvolve.
This procedure gives detections of diffuse emission from four
galaxies; UGC~1378, UGC~1455, UGC~1922 and UGC~2936; in all four cases 
the emission is associated with the galaxy center 
and mostly confined to the bulge (see Figure~3 which overlays
x-ray contours on the 2MASS K-band images). The emission is weakest 
for UGC~1455 which has only a small pool of diffuse gas. 

We made approximate estimates of the diffuse gas flux and luminosity using 
{\tt specextract}. The spectra were extracted from the galaxy centers and then 
examined using {\tt XSPEC}. The count statistics was poor and hence the fits
only approximate. But we were able to obtain a first estimate of the diffuse gas
luminosities in the centers of the galaxies (Table~1). For both UGC~1455
and UGC~2936 we detected a nuclear component (AGN) as well as diffuse emission. The
emission in UGC~2936 is clearly non-symmetric about
the galaxy center. Its origin may be disk star formation, or inverse
Compton emission from a radio lobe.  Deeper observations are required
to spectrally distinguish between the two possible origins.

\section{Discussion}

\noindent 
{\bf (i)~X-ray Bright AGN in LSB Galaxies~:}~The detection of x-ray
emission from the nuclei of UGC~2936 and UGC~1455 shows that though LSB
galaxies are metal poor and have little ongoing star formation, their
nuclei can host AGN activity that is bright in the x-ray domain. Another 
prominent LSB galaxy whose x-ray flux has been derived is UGC~6614 
(Naik, Paul \& Das, in preparation). It has an x-ray luminosity of 
$L_{x}\sim1.3\times10^{42}\,{\rm erg}\,{\rm s}^{-1}$ which is comparable to
that observed from UGC~2936 (Table~1). Thus the nuclear 
x-ray luminosity of UGC~1455 is comparable to that observed from the centers of
Low Luminosity AGNs but the x-ray luminosities of both UGC~2936 and UGC~6614 
are comparable to bright Seyfert nuclei. About 20~\% of LSB
galaxies show signs of AGN activity at optical wavelengths
\citep{Impey.etal.2001}, which is similar to late type spirals (Sc-Sm)
for which only about 15~\% show AGN activity \citep{Ho2008}. A large
fraction are also radio loud \citep{Das.etal.2006,Das.etal.2007}. In our sample,
excluding UGC~1455 and UGC~12845, the remaining galaxies are all radio loud 
(NVSS survey) \citep{Condon.etal.1998} and the morphology is often
a compact core with some associated extended emission representing
perhaps radio lobes or jets from the AGN (Das et al. 2008, in
preparation). Such a high fraction of radio cores are also detected in
Seyferts and LINERs \citep{Ho2008}. Thus AGN activity in LSB systems is
fairly similar to that seen in nearby Seyfert galaxies or LINERS, even
though the disk morphology and star formation rate is very different
from regular star forming galaxies on the Hubble Sequence.

\noindent
{\bf (ii)~Black Hole Masses~:}~The
clue to nuclear activity in giant LSB galaxies maybe the dominant
bulge that is often observed in these galaxies
\citep{Schombert.1998}. AGNs are more frequently found in bright galaxies
with large bulges \citep{Ho.Filippenko.Sargent.1997} and their 
formation and growth is linked
to the mass of the supermassive black hole $M_{BH}$ (SMBH)
\citep{Ferrarese.Ford.2005}. To get an idea of the black hole masses in LSB 
galaxies we applied the virial technique to the AGN line emission observed from these
galaxies. This method gives only an approximate estimate of $M_{BH}$ as the emission lines 
could be broadened by non-gravitational effects. The $M_{BH}$ can be derived
from the H$\alpha$ line luminositiy and linewidth \citep{Greene.Ho.2007}. 
Unfortunately, the relevant optical data is not available
for most LSB galaxies. Nor are these galaxies defined on the $M_{BH}-\sigma$ relation through 
other observations. Hence we were able to derive
$M_{BH}$ for only two galaxies: UGC~2936 and UGC~6614 using published H$\alpha$ line 
luminosities and linewidths \citep{Sprayberry.etal.1995,Schombert.1998,Kennicutt.etal.1984}. 
For UGC~2936 $M_{BH}=6.5\times10^{6}~M_{\odot}$ and for UGC~6614 $M_{BH}=2.9\times10^{7}~M_{\odot}$.
We then used the x-ray luminosities
(see Table~1 for UGC~2936 and for UGC~6614 $L_{x}\sim1.3\times10^{42}\,{\rm erg}\,{\rm s}^{-1}$) 
and the radio luminosities ($L_R$) of these galaxies  
to see where they lie on the $L_{X}$-$L_R$ fundamental plane \citep{Merloni.etal.2003}.
For UGC~2936 we used GMRT observations at 610~MHz and 1280~MHz to
determine the radio spectral index ($\alpha=0.55$) and then derived the  
$L_R$ at 5~GHz. Similarly for UGC~6614; however, in this galaxy the  
spectral index is flat and we used the flux density in the VLA 1.4~GHz map 
to derive $L_R$ at 5~GHz \citep{Das.etal.2006}. We find that both galaxies lie suprisingly 
well above the $L_{X}$-$L_R$ plane
and their $M_{BH}$ values are considerably lower than galaxies lying on the correlation as shown
in Figure~4. Thus though these LSB galaxies show AGN activity comparable to normal Seyferts, 
their nuclear black holes appear to be less massive than those detected in brighter galaxies.

\noindent
{\bf (iii)~Diffuse Emission from the Bulge~:}~This is the
first tentative detection of diffuse x-ray emission from LSB galaxies.
Such emission may arise from massive star forming regions and supernovae
\citep{Cui.etal.1996,Strickland.Heckman.2007} (e.g. M82), 
coronal x-ray emitting gas from galactic fountains
\citep{Fraternali.Binney.2008} or star formation in 
spiral arms \citep{Tyler.etal.2004} and is thus associated with 
metal enrichment in galaxies. In the
galaxies UGC~1378 and UGC~1922 the diffuse gas is mainly concentrated in 
the bulge (Figure~3). For UGC~1455 it is associated with the bulge and the
small, oval bar in the center, which may alternatively be a pseudobulge
\citep{deJong.1996}. In UGC~2936 the emission is concentrated in the bulge but also 
extended on one side (see (iv)). Since there is no apparent 
ongoing star formation activity in any of the galaxies except UGC~2936 
which shows patchy $H\alpha$ emission over the disk and nucleus 
\citep{Robitaille.etal.2007} the most likely origin for the diffuse
emission is the old stellar population in the bulges as 
well as AGN activity in the center of these galaxies. The luminosity is
$10^{36}$ to $10^{40}~erg~s^{-1}$ which is similar to that observed from the 
centers of nearby spiral galaxies \citep{Tyler.etal.2004}. We also determined the mid-infrared 
emission (at $12~\mu m$) for 5 galaxies in our sample; these 
were all IRAS values from NED (Table~1). The mid-IR flux values are similar to that 
observed from the centers of nearby bright galaxies in Tyler et al. (2004). 
This further supports the idea that the diffuse emission arises from star formation 
associated with the bulge.

\noindent
{\bf (iv)~Diffuse Emission Extending into the Disk in UGC~2936~:}~ The
diffuse emission in UGC~2936 clearly extends out into the disk on one side of the
galaxy center in the north-east direction (Figure~3). The simplest intrepretation
is spiral arm star formation. However, we do not detect similar emission from the other side.
The asymmetry suggests that we cannot rule out an x-ray
jet origin for the emission; the jet may be associated with the strong
AGN activity in the galaxy. Further observations are required to
understand this interesting feature.

\noindent
{\bf (v)~Bulge Evolution Independent of Disk Evolution ?~:}~Our observations 
show that LSB galaxies host AGN and relatively massive black holes 
despite having poorly evolved disks. They are similar to bulgeless late type 
spirals in their disk properties \citep{Boker.etal.2002} but closer to bulge dominated galaxies in 
their nuclear properties. Thus the disks and nuclei of these galaxies may
have evolved fairly independently of each other. One of the reasons for this 
kind of evolution could be the shape of the dark halo potential in LSB galaxies.
The halo potential is found to be 
relatively shallow in the center or bulge \citep{deNaray.etal.2008,Zackrisson.etal.2006}
but relatively strong in the disk. This would allow the center to evolve
whereas disk instabilities would be suppressed by the presence of the
dark matter halo at larger radii \citep{Mihos.etal.1997}. Alternatively, the bulges
may have evolved through other processes
such as galaxy collisisons \citep{Mapelli.etal.2008}; however the
disks in such processes have to remain fairly undisturbed which is a
tough constraint for such models. Another process could be the slow secular evolution 
of the disk through the formation of oval distortions or pseudobulges 
\citep{Kormendy.Kennicutt.2004}. 
The latter process is more likely as some giant LSB galaxies are found to have 
photometric signatures of pseudobulges \citep{Pizzella.etal.2008}. Overall these galaxies 
represent a good example of decoupled bulge-disk evolution and the underlying processes
should be investigated in more detail.

\section{CONCLUSIONS}

We present Chandra observations of eight giant, LSB galaxies all of which have 
a sizable bulge. Our main results are the following.

\noindent
{\bf (i)}~We have detected compact x-ray emission from the nuclei of two LSB galaxies, UGC~1455 
($L_{X}=1.1\times10^{40}~erg^{-s}$) and UGC~2936 ($L_{X}=1.8\times10^{42}~erg^{-s}$); it is due  
to AGN activity in these galaxies. The AGN emission is similar to that observed from
the centers of nearby Seyfert and LINER galaxies.

\noindent
{\bf (ii)}~For the galaxies UGC~2936 (our sample) and UGC~6614 (XMM archival data), we 
combined x-ray luminosities with radio luminosities and black hole masses to determine the 
location of these galaxies on the radio-x-ray fundamental plane. 
We find that both galaxies lie above the plane which suggests that 
their nuclei harbor less massive black holes compared to normal galaxies on the plane.

\noindent
{\bf (iii)}~Diffuse x-ray emission was detected from the bulges of four galaxies. The luminosity 
is similar to that observed from the centers of nearby star forming galaxies. These results 
combined with AGN emission suggests 
that the AGN and bulges of LSB galaxies have followed an evolutionary path similar to bulge dominated
bright galaxies even though their LSB disks are poorly evolved.

\noindent
{\bf (iv)}~The detection of AGN and diffuse emission from the bulges of LSB
galaxies shows that galaxies with unevolved disks can have normal bulges. Thus, whatever star
formation processes made the bulge did not make the disk component in these galaxies. In fact
the bulge and disk evolution appears to be distinctly decoupled in these galaxies. Giant LSB
galaxies are thus good sites to study decoupled bulge-disk evolutionary processes.

\acknowledgments
Support for this work was provided by the National Aeronautics and Space Administration through 
Chandra Award Number 08700344 issued by the Chandra X-ray Observatory Center, which is operated by 
the Smithsonian Astrophysical Observatory for and on behalf of the National Aeronautics Space 
Administration under contract NAS8-03060. This work has used XMM archival data on UGC~6614.
We have also used the 
NASA/IPAC Infrared Science Archive, which is operated by the Jet Propulsion Laboratory, 
California Institute of Technology, under contract with the National Aeronautics and Space Administration.
This publication makes use of data products from the Two Micron All Sky Survey, which is a joint project 
of the University of Massachusetts and the Infrared Processing and Analysis Center/California Institute of 
Technology, funded by the National Aeronautics and Space Administration and the National Science Foundation.
M.D. would also like to thank B.Paul and H.Raichur for useful discussions. We would also like to thank the
referee for useful comments.

\newpage 
\bibliographystyle{apj}

\bibliography{mdas.references}

\clearpage
{\scriptsize
\begin{deluxetable}{lcccccccc}
\rotate
\tablenum{1}
\tablewidth{0pt}
\tablecaption{Galaxy Sample and X-ray Fluxes}
\tablehead{
\colhead{Galaxy} & \colhead{Distance} & \colhead{Galaxy} & \colhead{Observation} & \colhead{Exposure} & \colhead{AGN} & \colhead{Diffuse Gas} & \colhead{Mid-IR} \\
\colhead{Name} & \colhead{(Mpc)} & \colhead{Position} & \colhead{ID} & \colhead{Time} & \colhead{Luminosity} & \colhead{Luminosity} & \colhead{Flux} \\
\colhead{} & \colhead{} & \colhead{RA, $\delta$ (J2000)} & \colhead{} & \colhead{(Ks)} & \colhead{(erg~s$^{-1}$)} & \colhead{(erg~s$^{-1}$)} & \colhead{(erg~cm$^{-2}$s$^{-1}$)}
}
\startdata
UGC~1455 & 67.3  & $01^{h}58^{m}48^{s}.0$, $+24^{\circ}53^{\prime}33^{\prime\prime}$ & 7764 & 3.79 & $1.1\times10^{40}$ & $5.5\times10^{39}$ & $<~2.0\times10^{-11}$ \\
UGC~2936 & 51.2  & $04^{h}02^{m}48{s}.2$, $+01^{\circ}57^{\prime}57^{\prime\prime}$ & 7769 & 2.74 & $1.8\times10^{42}$ & $7.5\times10^{40}$ & $7.3\times10^{-11}$ \\
\hline
UGC~1378 & 38.8 & $01^{h}56^{m}19^{s}.2$, $+73^{\circ}16^{\prime}58^{\prime\prime}$ & 7763 & 3.45 & $<~6.4\times10^{38}$ & $1.4\times10^{39}$ & $4.3\times10^{-11}$ \\
UGC~1922 & 150.0 & $02^{h}27^{m}45^{s}.8$, $+28^{\circ}12^{\prime}33^{\prime\prime}$ & 7884 & 5.88 & $<~5.7\times10^{39}$ & $1.4\times10^{40}$ & ... \\
UGC~3059 & 65.8 & $04^{h}29^{m}42^{s}.4$, $+03^{\circ}40^{\prime}55^{\prime\prime}$ & 7765 & 3.34 & $<~1.9\times10^{39}$ & ...& $1.8\times10^{-11}$ \\
UGC~4422 & 63.4 & $08^{h}27^{m}42^{s}.0$, $+21^{\circ}28^{\prime}44^{\prime\prime}$ & 7766 & 2.94 & $<~1.0\times10^{39}$ & ... & $<~2.7\times10^{-11}$\\
UGC~11754 & 62.6 & $21^{h}29^{m}31^{s}.5$, $+27^{\circ}19^{\prime}17^{\prime\prime}$ & 7767 & 4.15 & $<~7.0\times10^{38}$ & ... & ,,,\\
UGC~12845 & 63.9 & $23^{h}55^{m}41^{s}.9$, $+31^{\circ}53^{\prime}59^{\prime\prime}$ & 7768 & 3.25 & $<~1.8\times10^{39}$ & ... & ... \\

\enddata
\end{deluxetable}

\clearpage
\begin{figure}
\includegraphics[angle=0,scale=0.75]{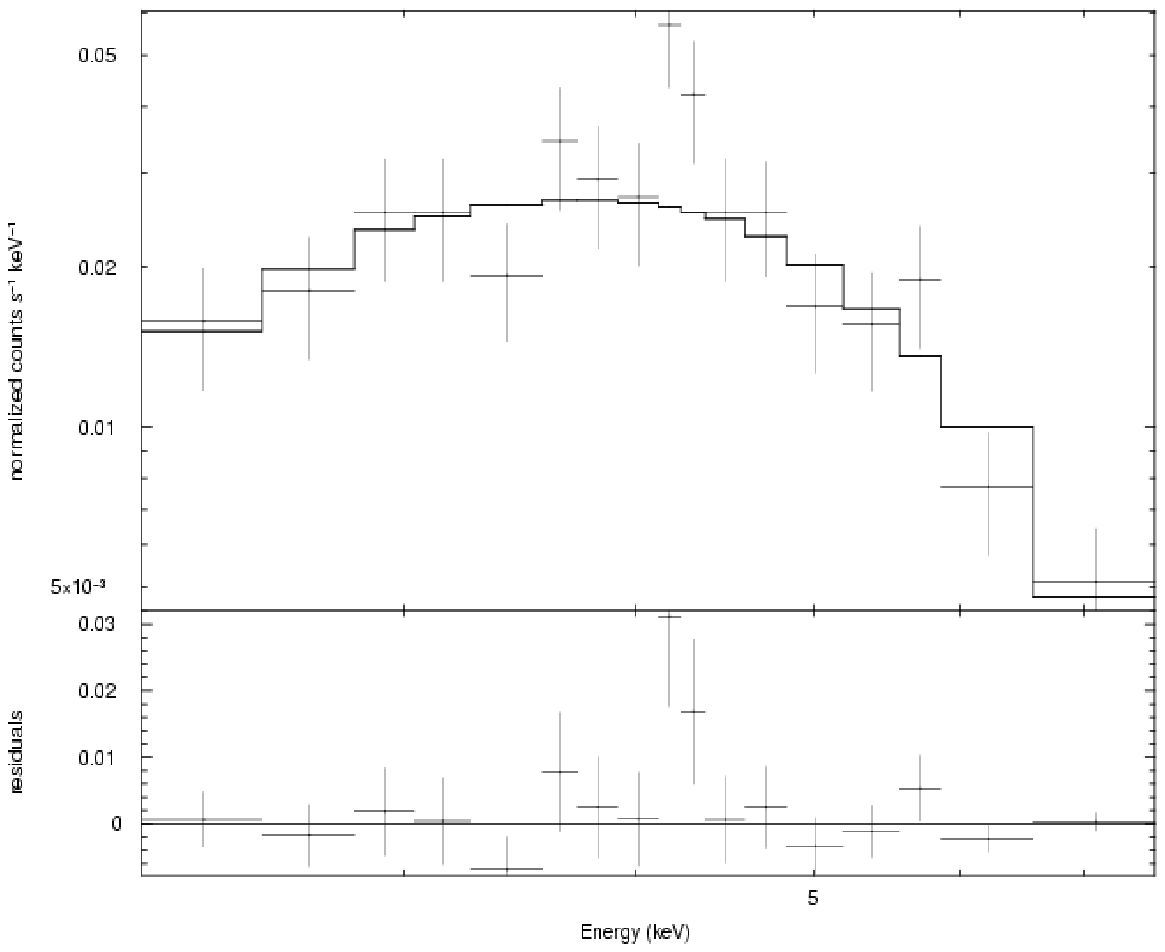}
\caption{X-ray spectrum of the emission from the nucleus of UGC~2936,
binned in the energy range 0.5--10\,keV. The corresponding fit is overlaid.}
\end{figure}

\clearpage
\begin{figure}
\includegraphics[angle=-90,scale=0.75]{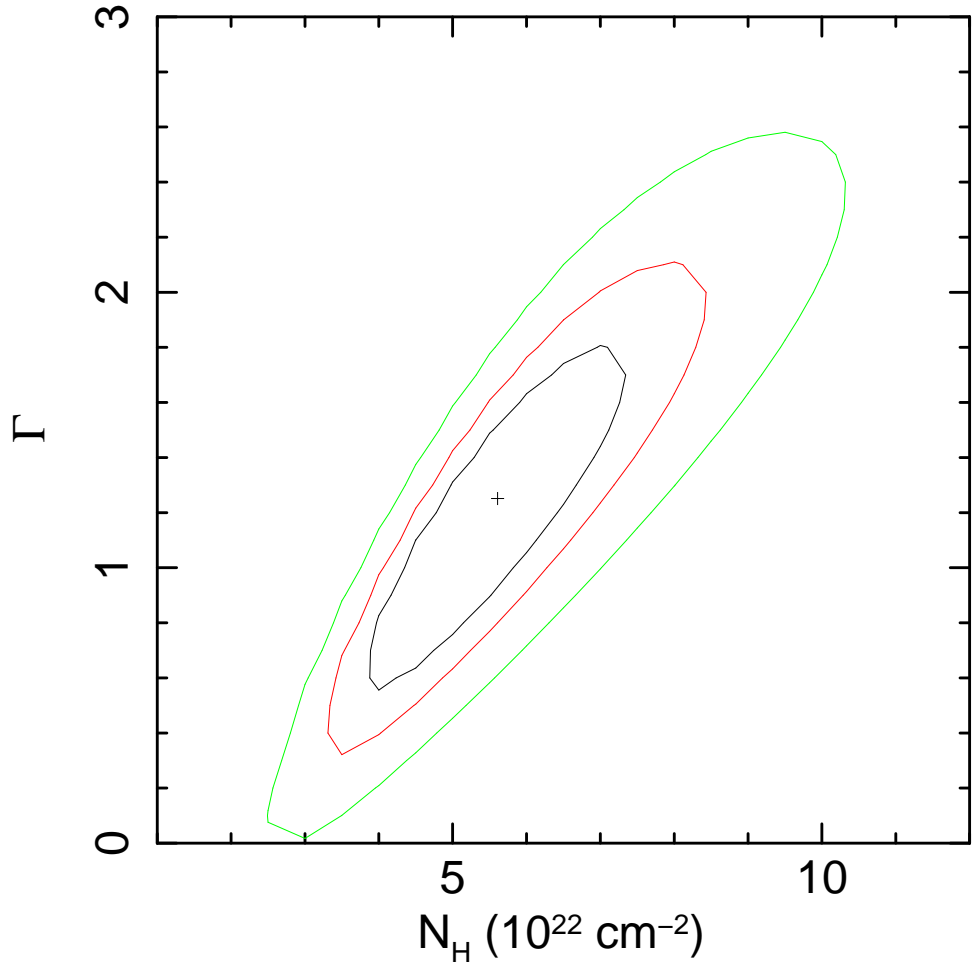}
\caption{Confidence contours (at the 68\%, 90\% and 99\% confidence
level for two interesting parameters) for the absorbed power-law fit
to the 0.5--10\,keV fit to UGC~2936.}
\end{figure}

\clearpage
\begin{figure}
\epsscale{0.35}
\plotone{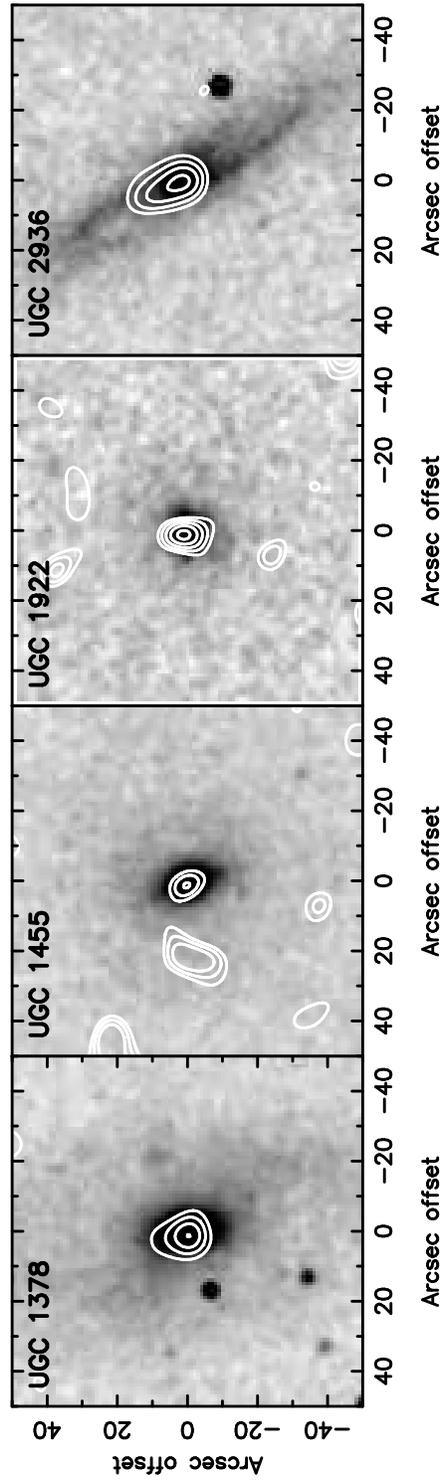}
\caption{Contours of diffuse emission superimposed over the 2MASS
near-infrared images of the galaxies UGC~1378, UGC~1922 and
UGC~2936. The contours are 8,9,10 and 11 times the noise level for UGC~1378, UGC~1922 
and UGC~2936. For UGC~1455 the contours are 9, 9.5 10}.
\end{figure}

\clearpage
\begin{figure}
\includegraphics[angle=-90,scale=0.75]{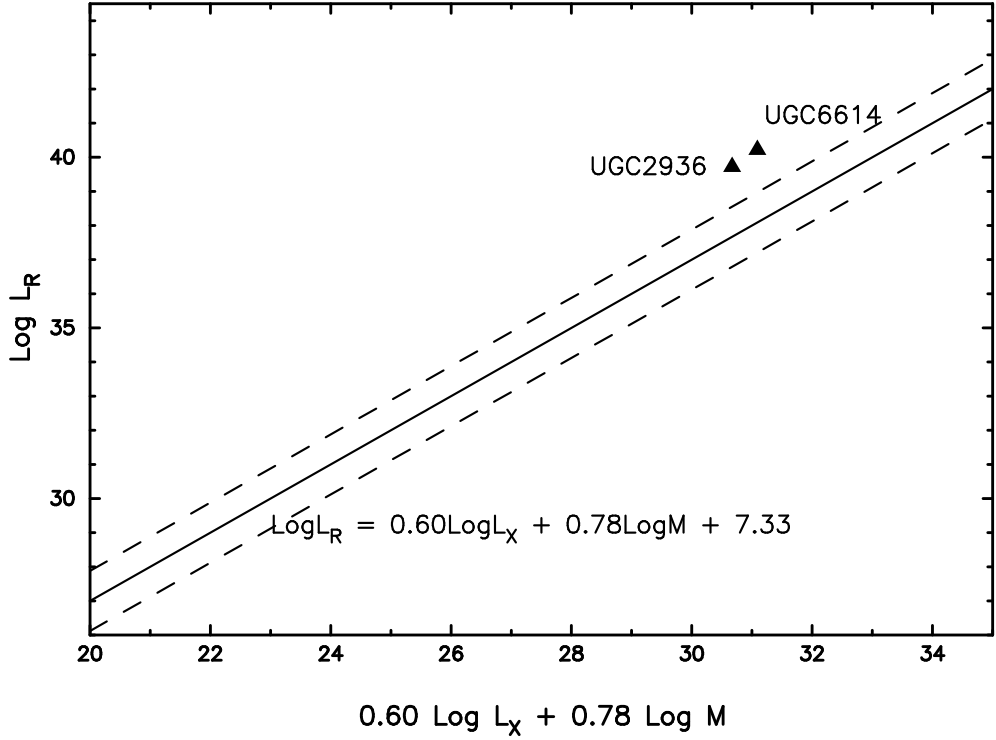}
\caption{Plot of the radio luminosity ($L_{R}$) against the x-ray luminosity ($L_{x}$) and black hole mass 
($M$). Both $L_{R}$ and $L_{x}$ are in $erg~s^{-1}$ and $M$ in solar units 
($M_{\odot}$). The solid line marks the fundamental plane of black hole activity and the dashed line is 
the approximate error width \citep{Merloni.etal.2003}. The errors for both UGC~2936 and UGC~6614 on either 
axes is much less than unity. Hence both galaxies lie well above the plane of black hole activity.
}
\end{figure}

\end{document}